\documentclass[reprint,showpacs,twocolumn,jcp,nobibnotes]{revtex4-1}
\usepackage{hyperref}
\hypersetup{colorlinks=true, citecolor=blue, urlcolor=blue, linkcolor=blue}
\usepackage{graphics,graphicx,amsfonts,amsmath,amsbsy,amssymb,color}
\usepackage{bm}
\usepackage{subfigure}
\usepackage{threeparttable}
\usepackage{amsmath}

\begin{document}

\title{%
Particle-particle ladder based basis-set corrections applied to
atoms and molecules using coupled-cluster theory
}

\author{Andreas Irmler}
\author{Andreas Gr\"uneis}
\email{andreas.grueneis@tuwien.ac.at}
\affiliation{
  Institute for Theoretical Physics, TU Wien,\\
  Wiedner Hauptstra\ss e 8-10/136, 1040 Vienna, Austria
}

\begin{abstract}
We investigate the basis-set convergence of electronic correlation energies
calculated using coupled cluster theory
and a recently proposed finite basis-set correction technique.
The correction is applied to atomic and molecular systems and
is based on a diagrammatically decomposed coupled cluster singles and doubles correlation energy.
Only the second-order energy and the
particle-particle ladder term are corrected for their basis-set incompleteness error.
We present absolute correlation energies and results for a large benchmark set.
Our findings indicate that basis set reductions by two cardinal numbers are possible
for atomization energies, ionization potentials and electron affinities without
compromising accuracy when compared to conventional CCSD calculations.
In the case of reaction energies we find that reductions by one cardinal number are possible
compared to conventional CCSD calculations.
The employed technique can readily be applied to other many-electron
theories without the need for three- or four-electron integrals.
\end{abstract}
\date{\today, Accepted manuscript}
\pacs{}

\keywords{}
\maketitle

\section{\label{sec:introduction}Introduction}

Quantum chemical many-electron theories that employ
anti-symmetrized one-particle functions (Slater determinants) as a basis for
the many-electron wavefunction
exhibit a frustratingly slow convergence of central quantities like the ground state
energy to the complete basis-set limit.
A large fraction of the computational cost involved in many-electron
theory calculations of atoms, molecules and solids originates from
the need to include large numbers of one-electron basis functions
necessary to achieve the desired level of precision.
Many techniques have been developed to accelerate the convergence
to the complete basis-set limit including
explicitly correlated methods, transcorrelated methods
or simple yet less efficient basis-set extrapolation
techniques~\cite{Hattig:CR112-4,Tenno2017,Knizia2009,Boys1969a,Boys1969b,TENNO2000169,Feller2013,PhysRevB.86.035111}.
More recently a density-functional theory based approach has also been employed to correct
for basis set incompleteness errors~\cite{Giner2018}.

It has been known since the early days of electronic structure theory that the
slow convergence of the wavefunction expansion in
Slater determinants is due to short-ranged interelectronic correlation.
As the electrons coalesce, a derivative discontinuity or `cusp' must arise,
so that a divergence in the kinetic energy operator cancels an opposite one in
the potential~\cite{Kato1957,Pack:JCP45-556,Ashcroft82,Sahni2003}.
Explicitly correlated methods account for
the cusp condition in an \emph{a priori} manner and are commonly referred to as
F12 theories, where F12 stands for a two-electron correlation factor
that enables a compact expansion of the wavefunction at short
interelectronic distances~\cite{Hattig:CR112-4,Werner:JCP126-164102,Kutzelnigg:JCP94-1985}.
F12 theories introduce, however, the need for additional many-electron
integrals such as three- and sometimes even four-electron integrals.
Due to their large computational overhead, explicitly correlated methods are
generally only beneficial for more complex parent methods. Furthermore they cause
substantial additional effort in their computer code implementation.
Nonetheless, thanks to their improved basis-set convergence and reliability,
they have become an indispensable tool for quantum
chemical calculations of large systems~\cite{Knizia2009,Hattig2010-ql,
Hattig:CR112-4,Kong:CR112-75,Tew:BOOK2010,Valeev2012,Werner:BOOK2010,Tenno12,Tenno12_2,Shiozaki2010}.

In this work we apply a recently proposed basis-set extrapolation
technique for coupled cluster singles and doubles (CCSD) theory to
atoms and molecules.
The employed technique is based on a diagrammatic
decomposition of the coupled cluster correlation energy and was previously
employed to study the uniform electron gas~\cite{Irmler2019a}.
An analysis of the decomposed correlation energy for large basis sets
in the uniform electron gas shows that the slow convergence of the
CCSD energy originates from the second-order correlation energy and
the so-called particle-particle ladder term.
Both contributions converge at the same rate but with opposite sign to
the complete basis-set (CBS) limit.
As a consequence MP2 theory exhibits a larger basis-set incompleteness
error than CCSD theory.
This observation is in agreement with prior findings for molecules and atoms where
it has been referred to as \emph{interference} effect~\cite{Petersson1988,Nyden1980,Petersson1981}.
Furthermore  related findings have been discussed in the context of
explicitly correlated methods~\cite{Kutzelnigg:JCP94-1985,Noga:JCP101-7738,Vogiatzis2011}.
We stress that the asymptotic convergence of the ppl contribution
has already been studied by Klopper and Kutzelnigg using a partial wave expansion in
third-order perturbation theory in Ref.~\cite{Kutzelnigg:JCP94-1985}.



Based on the observation of \emph{only two} dominant channels in the basis-set limit,
we have devised a basis-set correction
that is based on re-scaling the partice-particle ladder (ppl) correlation
energy contribution by a factor that is estimated from the corresponding basis-set
incompleteness error of the MP2 correlation energy, which can be obtained
in a computationally significantly cheaper manner albeit requiring a
non-negligible effort~\cite{Liakos2013}. Previously this scheme
has successfully been applied to estimate the CCSD correlation energy of
semiconducting and insulating solids~\cite{Irmler2019a}.
In this article we investigate the performance of the technique for a large set
of molecules. We study total energies, reaction energies, atomization energies,
ionization potentials, and electron affinities of a large benchmark set taken from
Ref.~\cite{Knizia2009}.
We stress that the presented scheme is related to the CCSD-F12a approximation, which
corrects for the basis-set incompleteness errors in the same terms, albeit using a more
sophisticated explicitly correlated ansatz~\cite{Knizia2007}.
Furthermore we note in passing that a similar rescaling approach of the
perturbative triples electronic correlation energy contribution has also been applied
in Ref.~\cite{Knizia2009}.

We emphasize that in addition to explicitly correlated methods, there exist a wide range
of basis set extrapolation techniques that can also be used to rapidly converge correlation
energies to the complete basis set limit~\cite{Feller2013,Ranasinghe2013}.
However, in this work we
compare to explicitly correlated methods that are widely-used for the prediction of
reliable CBS limit reference values. For the sake of consistency we employ the same
molecular geometries as employed for CCSD-F12a and CCSD-F12b published in Ref.~\cite{Knizia2009}.

\section{\label{sec:theory} Theory}


Here, we employ a recently proposed method to correct for the
basis-set incompleteness error of CCSD correlation energies.
The central premise of this method is that the CCSD correlation energy,
\begin{equation}
E^{\rm CCSD}_{\rm c}= 
	T_{ij}^{ab} \left( 2 \left<ij\middle|ab\right> - \left<ji\middle|ab\right> \right),
\label{eq:Ecorr3}
\end{equation}
can be decomposed into different diagrammatic contributions such that~\cite{Irmler2019a}
\begin{equation}
E_{\rm c}^{\rm CCSD}=E^{\rm driver}+  E^{\rm ppl}+ \underbrace{ E^{\rm phl}+ E^{\rm hhl}+ E^{\rm phr}+ ... }_{=E^{\rm rest}} ,
\label{eq:Ecorr}
\end{equation}
where $E^{\rm driver}$ corresponds to the MP2 correlation energy
\begin{equation}
E^{\rm driver}=  W_{ij}^{ab} \left \langle ab | ij \right \rangle
\label{eq:Ecorr2}
\end{equation}
and the particle-particle ladder term is defined as
\begin{equation}
E^{\rm ppl}= 
W_{ij}^{ab}
 \left \langle ab | cd \right \rangle
T_{ij}^{cd}.
\label{eq:EcorrPPL}
\end{equation}
$T_{ij}^{cd}$ is computed from the CCSD singles ($t_i^a$)
and doubles ($t_{ij}^{ab}$) amplitudes such that
$T_{ij}^{ab}=t_{ij}^{ab}+t_i^a t_j^b$.
$t_i^a$ and $t_{ij}^{ab}$ are obtained by solving the
corresponding amplitude equations\cite{Bartlett07,Cizek71}.
$W_{ij}^{ab}$ is given by
\begin{equation}
W_{ij}^{ab}= \frac{
2 \left \langle ij | ab \right \rangle- \left \langle ji | ab \right \rangle }
{\epsilon_i + \epsilon_j - \epsilon_a - \epsilon_b}.
\label{eq:Wijab}
\end{equation}
The employed indices are explained and summarized in Table~\ref{tab:index}.
Einstein summation convention applies to repeated indices
throughout this work.

\begin{table}[t]
\caption{
Index notation for different orbital subspaces of the complete one-electron basis. 
The orbitals refer to the Hartree-Fock spatial orbital basis set
over which the wavefunction amplitudes are defined.
}
\label{tab:index}
\begin{ruledtabular}
\begin{tabular}{lccc}
                        &  Occ. orbitals & Virt. orbitals & Complement \\ \hline
$i,j,k,l,m,n$           &  Yes       &  No  & No \\
$a,b,c,d$               &   No       &  Yes & No \\
$C$                     &   No       &  Yes & Yes \\
$P,Q$                   &  Yes       &  Yes & Yes \\
\end{tabular}
\end{ruledtabular}
\end{table}


We note that the employed decomposition of the CCSD correlation energy
is achieved by replacing the doubles amplitudes in Eq.~(\ref{eq:Ecorr3})
with the corresponding right hand side
of the amplitude equations using converged CCSD amplitudes.
For the sake of brevity, we define $E^{\rm rest}$ such that it contains all remaining
contributions to the CCSD correlation energy including terms such as
the particle-hole ladder, hole-hole ladder and the particle-hole
ring.
We stress that the labelling of the contributions to the correlation energy
on the right hand side of Eq.~(\ref{eq:Ecorr}) is inspired by the corresponding
terms in the amplitude equations and does not imply that the ppl term includes particle-particle
ladder contributions to the correlation energy only.
The latter holds only in the case
of $t_{ij}^{ab}={t^{(1)}}_{ij}^{ab}$ where
${t^{(1)}}_{ij}^{ab}=
\frac{\left \langle ab | ij \right \rangle}
{\epsilon_i + \epsilon_j - \epsilon_a - \epsilon_b}$.
We note that a similar labeling of the terms in the amplitude equations
is used by Shepherd et al. in Ref.~\cite{Shepherd2014}.
We refer the reader to Ref.~\cite{Irmler2019a} for a more
detailed analysis of the different diagrammatic contributions to the CCSD
correlation energy in periodic systems. 

The MP2 correlation energy, $E^{\rm driver}$, converges as $1/L^3$ 
using a correlation consistent atom-centered Gaussian basis set, 
where $L$ refers to the cardinal number of the employed basis set.
We will argue in the following that the first two terms on the right hand
side of Eq.~(\ref{eq:Ecorr})
exhibit the same convergence rate when approaching the complete basis set limit,
constituting the dominant source of the basis set incompleteness error of
the CCSD correlation energy.
To better understand the convergence behaviour of $E^{\rm ppl}$,
we approximate the first-order amplitudes using the following expression that
follows from F12 theory
\begin{equation}
{t^{(1)}}_{ij}^{ab} \approx \left \langle ab
\left | (-1)\frac{t_{ij}}{\gamma} e^{-\gamma r_{12}} \right | ij \right \rangle.
\label{eq:texp12}
\end{equation}
$t_{ij}$ are geminal amplitudes determined by the Kato cusp conditions~\cite{Kato1957}.
The expression above implies that the first-order doubles amplitudes
can be approximated using an
expression that is similar to the corresponding electron repulsion integral,
where the Coulomb kernel has been replaced by the Slater-type correlation factor that
depends on a parameter $\gamma$.
Approximating the CCSD doubles amplitudes $t_{ij}^{ab}$ on the right hand side of
Eq.~(\ref{eq:EcorrPPL}) by Eq.~(\ref{eq:texp12})
and disregarding the contribution of single amplitudes,
we can approximate the particle-particle ladder
contribution to the correlation energy by
\begin{equation}
E^{\rm ppl}\approx 
W_{ij}^{ab}
 \left \langle ab | cd \right \rangle
 \left \langle cd \left | (-1)\frac{t_{ij}}{\gamma} e^{-\gamma r_{12}}
\right  | ij \right \rangle
\label{eq:EcorrPPLapprox}
\end{equation}
We now replace the summation over the virtual orbital
indices $c$ and $d$ in Eq.~(\ref{eq:EcorrPPLapprox}) by the closure relation
$
|c \rangle \langle c |  = |P \rangle \langle P |
                          - |k \rangle \langle k | $,
yielding the following approximate expression for $E^{\rm ppl}$
\begin{widetext}
\begin{equation}
E^{\rm ppl}  \approx
W_{ij}^{ab}(-1)\frac{t_{ij}}{\gamma}
\left [
 \left \langle ab \left | \frac{e^{-\gamma r_{12}}}{r_{12}} \right | ij \right \rangle
\underbrace{+
\left \langle ab | kl \right \rangle
\left \langle kl | e^{-\gamma r_{12}} | ij \right \rangle
 -
\left \langle ab | Pl \right \rangle
\left \langle Pl | e^{-\gamma r_{12}} | ij \right \rangle
}_{= - \left \langle ab | Cl \right \rangle
\left \langle Cl | e^{-\gamma r_{12}} | ij \right \rangle
}
-
\left \langle ab | kQ \right \rangle
\left \langle kQ | e^{-\gamma r_{12}} | ij \right \rangle
\right ]
\label{eq:EcorrPPLapprox2}
\end{equation}
\end{widetext}
We stress that the particle-particle ladder contribution to the correlation
energy is positive (for $t_{ij}^{ab} \approx {t^{(1)}}_{ij}^{ab}$),
whereas the last three terms in the brackets on the right hand side
of Eq.~(\ref{eq:EcorrPPLapprox2}) yield a negative contribution.
According to this, the leading contribution to the correlation energy must be
the first term in the brackets on the right hand side of the above expression.
This term converges at the same rate to the complete basis set limit as $E^{\rm driver}$.
In particular the
$\frac{e^{-\gamma r_{12}}}{r_{12}}$
kernel exhibits a singularity at $r_{12}=0$ that causes an asymptotic
basis set convergence rate which is identical to MP2 theory ($1/L^3$).
In short, we have shown above that $E^{\rm ppl}$ defined by Eq.~(\ref{eq:EcorrPPL})
will converge to the CBS limit at the same rate as $E^{\rm driver}$ if the
following approximations hold
${t}_{ij}^{ab} \approx {t^{(1)}}_{ij}^{ab} \approx \left \langle ab
\left | (-1) \frac{t_{ij}}{\gamma} e^{-\gamma r_{12}} \right  | ij \right \rangle$.

\begin{figure*}[t]
\begin{center}
\includegraphics[width=0.99\linewidth]{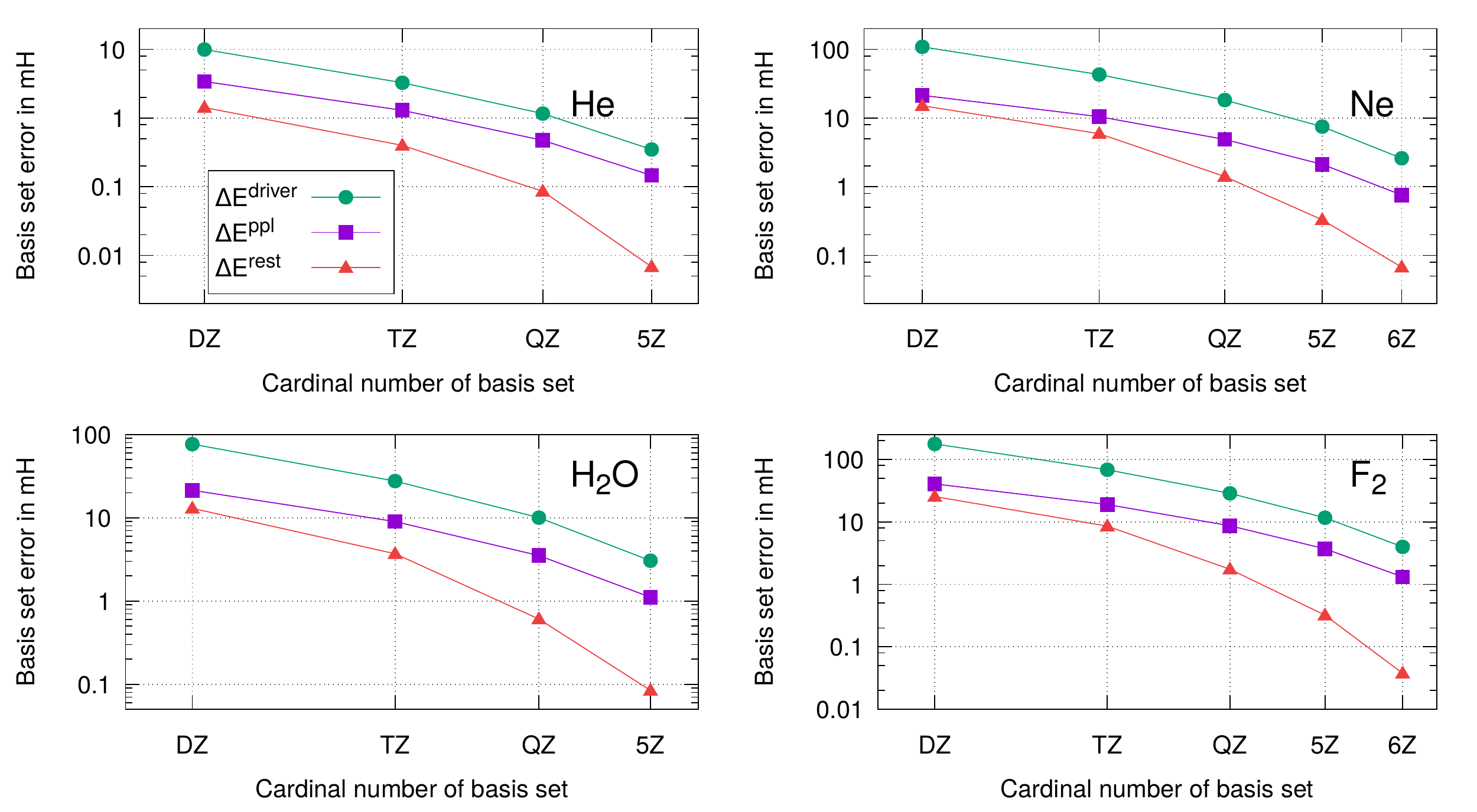}
\end{center}
\caption{\label{fig:channel}Absolute basis set error
          of CCSD valence electron correlation energy contributions
          ($\Delta E^{\rm driver}(X)$, $\Delta E^{\rm ppl}(X)$ and $\Delta E^{\rm rest}(X)$)
          for He,  Ne, H$_2$O and F$_2$ using the AVXZ basis sets.
	  The CBS limit reference energy is approximated using AV6Z for He and H$_2$O, wheras AV7Z has been
          used for Ne and F$_2$.
 	}
\end{figure*}

We now put the analysis carried out above to a numerical test.
Figure~\ref{fig:channel} depicts the convergence of the absolute basis set incompleteness
error in the decomposed correlation energy contributions of He, Ne, H$_2$O and F$_2$.
The errors are retrieved as a function of the cardinal number of the employed aug-cc-pV$X$Z basis.
Basis set incompleteness errors are defined by
\begin{equation*}
\Delta E(X)= E(\rm CBS) - E(X).
\end{equation*}
We refer to channel decomposed correlation energies obtained in a given basis set $X$ by
$E^{\rm driver}(X)$, $E^{\rm ppl}(X)$ and $E^{\rm rest}(X)$.
For the results shown in Figure~\ref{fig:channel}, we estimate the CBS values using
an aug-cc-pV6Z basis for He and H$_2$O, and an aug-cc-pV7Z basis for Ne and F$_2$.
We note that $\Delta E^{\rm driver} (X)$ and $\Delta E^{\rm ppl} (X)$
are parallel on a logarithmic scale, whereas
$\Delta E^{\rm rest}(X)$ exhibits a faster convergence rate compared to the other terms.
From these results and the analysis carried out above as well as for the uniform
electron gas~\cite{Irmler2019a}, we conclude that $E^{\rm driver}$ and $E^{\rm ppl}$
are the \emph{only} two dominant channels in the basis-set limit of coupled-cluster
singles and doubles theory.
Furthermore we do not identify a fundamental difference in the discussed behaviour
between any of these systems. 



Based on the analysis above and
in agreement with Ref.~\cite{Irmler2019a},
it is reasonable to propose a finite basis set error
correction to the CCSD correlation energy that
requires the CBS limit estimate of the second-order correlation
energy $E^{\rm driver}({\rm CBS})$ only.
We define the approximation to the CBS limit estimate of the CCSD correlation energy,
CCSD-PPL, as
\begin{equation}
E^{\rm CCSD-PPL}_{\rm c} =
E^{\rm CCSD}_{\rm c}(X)+
\Delta E^{\rm driver}(X)+ \Delta E^{\rm ppl}(X),
\label{eq:CCSDPPL}
\end{equation}
where we employ the following approximation to the particle-particle ladder correlation energy contribution
\begin{equation}
\Delta E^{\rm ppl}(X) = \frac{E^{\rm driver}(\rm CBS)}{E^{\rm driver}(X)} E^{\rm ppl}(X)  - E^{\rm ppl}(X)
\label{eq:DPPL}
\end{equation}

\section{\label{sec:comp}Computational Details}

We have modified the coupled cluster code in the
open-source quantum chemistry package $\texttt{PSI4}$~\cite{PSI4ref}
such that the $E^{\rm ppl}(X)$ contribution to the CCSD correlation
energy is computed separately at the end of each CCSD calculation
using fully converged CCSD amplitudes.
The required modifications to the exisiting CCSD code are minor
and we expect that other computer code implementations
of coupled cluster theory can also be modified in such a simple manner.
Throughout this work, correlation energies always correspond to
valence electron correlation energies only.

We employ aug-cc-pVXZ basis sets for first-row
elements and aug-cc-pV(X+d)Z basis sets for second-row elements.
These basis sets will be denoted as AVXZ throughout the article.
For reaction energies (REs), atomization energies (AEs), ionization
potentials (IPs), and electron affinities (EAs) CBS limit estimates
are obtained from AVQZ and AV5Z CCSD correlation energies using the
$E_n=E_\textrm{CBS} + a/n^3$ extrapolation formula and will be denoted as CCSD/[45].
Likewise the MP2 CBS limit estimates, needed for CCSD-PPL calculations,
are also obtained using a [45] extrapolation.
The employed equilibrium geometries have been taken from Ref.~\cite{Knizia2009}.


\section{\label{sec:results}Results}

\begin{table}
\caption{
Valence electron correlation energies for H$_2$O and F$_2$ in mH.
The CCSD CBS limit values are -297.9~mH and -601.17~mH, respectively~\cite{Knizia2009}.
}
\label{tab:ecorrval}
\begin{ruledtabular}
\begin{tabular}{l ccc}
                               \multicolumn{4}{c}{H$_2$O}      \\ \hline
                              &    AVDZ &    AVTZ &    AVQZ     \\ \hline
CCSD                          & -227.11 & -273.05 & -288.21     \\
CCSD-PPL                      & -291.44 & -298.22 & -299.50     \\
CCSD-F12a~\cite{Knizia2009}   & -293.22 & -298.50 & -299.37     \\
CCSD-F12~\cite{Shiozaki2008}  & -289.86 & -295.40 & -297.23     \\
\hline
                               \multicolumn{4}{c}{F$_2$}    \\ \hline
                              &   AVDZ & AVTZ & AVQZ         \\  \hline
CCSD                          & -435.39 & -538.91 & -575.10  \\
CCSD-PPL                      & -580.92 & -598.32 & -602.67  \\
CCSD-F12a~\cite{Knizia2009}   & -590.81 & -599.65 & -602.89  \\
CCSD-F12~\cite{Shiozaki2008}  & -584.83 & -594.98 & -599.01  \\

\end{tabular}
\end{ruledtabular}
\end{table}

\subsection{Total energies}
\label{sec:totenergies}
We first seek to assess the efficiency of CCSD-PPL
for total valence electron correlation energies of the closed
shell molecules H$_2$O and F$_2$.
Table~\ref{tab:ecorrval} summarizes the calculated valence electron correlation
energies for AVXZ  basis sets with $X=\mathrm{D,T,Q}$ and results
obtained using F12 approaches
taken from Refs.~\cite{Knizia2009,Shiozaki2008}
The valence electron correlation energies shown in
Table~\ref{tab:ecorrval} demonstrate that CCSD-PPL
converges to the CBS limit with respect to the employed basis set
at a similar rate as F12 theories.
Compared to CCSD theory, CCSD-PPL allows for a reduction by
approximately two cardinal numbers in the employed basis set size
while capturing a similar fraction of the correlation energy.\\
A careful analysis of the energy contributions from the different channels shows for H$_2$O
that $|\Delta E^{\rm rest}|$ is 12.8~mH and 3.5~mH for AVDZ and AVTZ, respectively.
These numbers are even larger than the total deviations of the CCSD-PPL results
compared to the CBS limit (see Table~\ref{tab:ecorrval}).
For F$_2$ we observe a similar behaviour.
Therefore we conclude that the good performance of CCSD-PPL for total correlation
energies using small basis sets, such as AVDZ, is partly due to error cancelation between
$\Delta E^{\rm rest}$ and $\Delta E^{\rm ppl}$.

\subsection{Reaction energies}

To further assess the basis set convergence of CCSD-PPL, we calculate reaction energies using
a benchmark set developed by Knizia et al.~\cite{Knizia2009}. 
However, instead of the full set containing 54 closed-shell molecules, and 50
open-shell molecules, we restrict our analysis to a subset of 28 reactions, containing small closed-shell
molecules only. The investigated reactions are listed in Table~\ref{tab:listreactions}.

\begin{figure*}[t]
\includegraphics[width=18.0cm,clip=true]{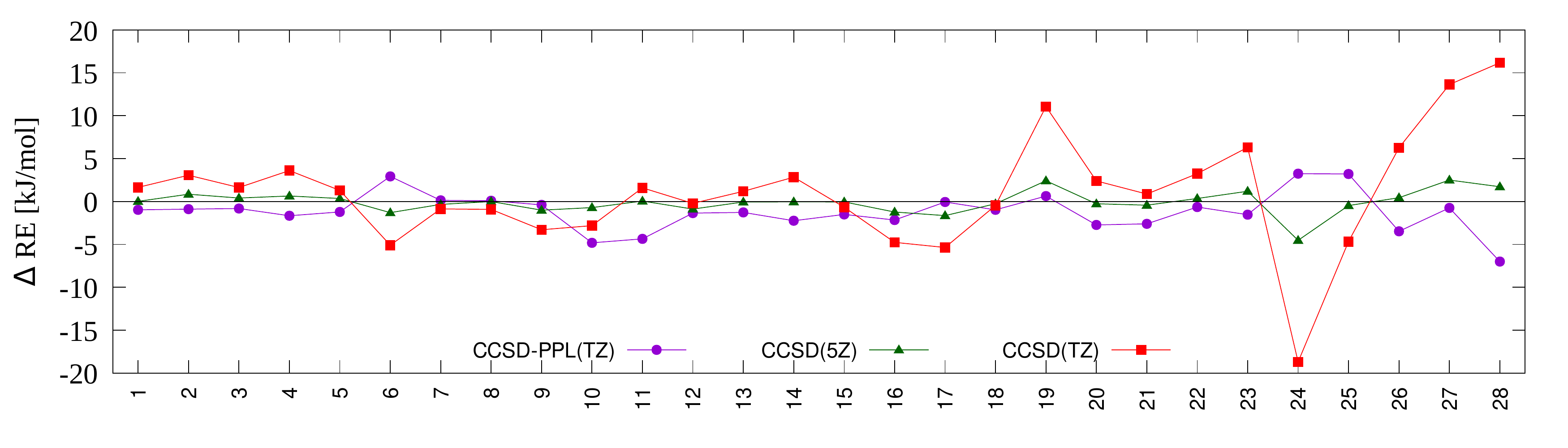}
\caption{\label{fig:reaction}
	  Basis set incompleteness errors of computed correlation energy
          contributions to reaction energies.
          The investigated reactions are listed in Table~\ref{tab:listreactions}.
          The required CBS limit results have been approximated using CCSD/[45].
}
\end{figure*}

\begin{table}[ht]
\caption{
	Statistical analysis of basis set incompleteness errors of correlation energy
        contributions to the reaction energies.
        Mean absolute deviations (MAD), root mean squared (RMS) and maximum deviations (MAX)
        have been computed for all reactions listed in Table~\ref{tab:listreactions}.
        The CBS limit reference results have been approximated using CCSD/[45].
        F12a/b values were extracted from the supplemental information from Ref.~\cite{Knizia2009}.
        Given F12a/b values are total valence electron energies including RHF contribution with CABS 
	singles corrections.
}
\label{tab:statreactions}
\begin{ruledtabular}
\begin{tabular}{l ccc }
	              & MAD    &   RMS   & MAX   \\ \hline
CCSD/AVDZ            & 10.906 & 13.837 & 43.053 \\
CCSD/AVTZ            &  4.451 &  6.485 & 18.707 \\
CCSD/AVQZ            &  1.674 &  2.545 &  8.838 \\
CCSD/AV5Z            &  0.863 &  1.305 &  4.532 \\
\\
CCSD-PPL/AVDZ         &  5.908 &  9.231 & 35.917 \\
CCSD-PPL/AVTZ         &  1.918 &  2.493 &  7.005 \\
CCSD-PPL/AVQZ         &  1.070 &  1.310 &  2.852 \\
\\
CCSD-F12a/AVTZ\cite{Knizia2009}        &  1.047 & 1.392 &  4.625 \\
CCSD-F12a/AVQZ\cite{Knizia2009}        &  0.395 & 0.507 &  1.145 \\
\\
CCSD-F12b/AVTZ\cite{Knizia2009}        &  0.891 &  1.392 &  5.313 \\
CCSD-F12b/AVQZ\cite{Knizia2009}        &  0.457 &  0.585 &  1.304 \\

\end{tabular}
\end{ruledtabular}
\end{table}

Figure~\ref{fig:reaction} depicts basis set incompleteness errors of the computed
correlation energy contributions to the reaction energies for CCSD/AVTZ, CCSD/AV5Z and
CCSD-PPL/AVTZ. We note that the errors of CCSD/AVTZ results are significantly larger than corresponding
CCSD-PPL/AVTZ results, whereas CCSD/AV5Z exihibits smaller errors than CCSD-PPL/AVTZ.
This indicates that in contrast to total correlation energies,
CCSD-PPL does not allow for a reduction by two cardinal numbers in the basis compared to CCSD,
while capturing the same fraction of the CBS limit correlation energy contribution to
the reaction energies.
This observation is also reflected in the statistical analysis of the computed results summarized in
Table~\ref{tab:statreactions}. 
CCSD/AVXZ is closer to the CBS limit than CCSD-PPL/AV(X-2)Z for RMS, MAD and MAX.
However, CCSD-PPL/AV(X-1)Z and CCSD/AVXZ exhibit similar basis set incompleteness errors, indicating
the a reduction by one cardinal number can be achieved using CCSD-PPL for reaction
energies.
By comparing CCSD/AVXZ to CCSD-PPL/AVXZ, we find that the RMS, MAD and MAX of basis set incompleteness errors
are reduced by approximately a factor of two, except for AVDZ, where this factor is smaller. 
However, this could be related to non-negligible basis set incompleteness errors of the underlying
HF reference.
In contrast to CCSD-PPL, both F12a and F12b approximations allow for a reduction by two
cardinal numbers while achieving a similar level of accuracy as corresponding CCSD results.
In particular, F12a and F12b achieves CCSD/AV5Z quality using AVTZ basis sets only.
For a comparable accuracy the PPL scheme requires AVQZ basis sets.


\begin{table*}
\small
\caption{
        List of closed shell reactions.
}
\label{tab:listreactions}
\begin{ruledtabular}
\begin{tabular}{l l l l l l }
NR &  Reaction    &  NR &  Reaction  &  NR &  Reaction      \\ \hline
1  &  CO + H2 $\rightarrow$ HCHO                  &
2  &  CO + H2O $\rightarrow$  CO2 + H2            &
3  &  CH3OH + HCl $\rightarrow$  CH3Cl + H2O      \\
4  &  H2O + CO  $\rightarrow$ HCOOH               &
5  &  CH3OH + H2S  $\rightarrow$ CH3SH + H2O      &
6  &  CS2 + 2 H2O  $\rightarrow$ CO2 + 2 H2S      \\
7  &  C2H6 + H2  $\rightarrow$ 2 CH4              &
8  &  HNCO + H2O $\rightarrow$  CO2 + NH3         &
9  &  CH4 + Cl2 $\rightarrow$  CH3Cl + HCl        \\
10 &  Cl2 + F2 $\rightarrow$  2 ClF               &
11 &  CO + Cl2 $\rightarrow$  COCl2               &
12 &  CO2 + 3 H2 $\rightarrow$  CH3OH + H2O       \\
13 &  HCHO + H2 $\rightarrow$  CH3OH              &
14 &  CO + 2 H2 $\rightarrow$  CH3OH              &
15 &  C2H4 + H2 $\rightarrow$  C2H6               \\
16 &  SO3 + CO $\rightarrow$  SO2 + CO2           &
17 &  H2 + Cl2 $\rightarrow$  2 HCl               &
18 &  C2H2 + H2 $\rightarrow$  C2H4               \\
19 &  SO2 + H2O2 $\rightarrow$  SO3 + H2O         &
20 &  CO + 3 H2 $\rightarrow$  CH4 + H2O          &
21 &  HCN + 3 H2 $\rightarrow$  CH4 + NH3         \\
22 &  H2O2 + H2 $\rightarrow$  2 H2O              &
23 &  CO + H2O2 $\rightarrow$  CO2 + H2O          &
24 &  2 NH3 + 3 Cl2 $\rightarrow$  N2 + 6 HCl     \\
25 &  3 N2H4 $\rightarrow$  4 NH3 + N2            &
26 &  H2 + F2 $\rightarrow$  2 HF                 &
27 &  CH4 + 4 H2O2 $\rightarrow$  CO2 + 6 H2O     \\
28 &  2 NH3 + 3 F2 $\rightarrow$  N2 + 6 HF       & & \\
\end{tabular}
\end{ruledtabular}
\end{table*}


\subsection{Atomization energies, ionization potentials, and electron affinities}

\begin{table*}
\caption{
	Statistical analysis of basis set incompleteness errors of computed
        atomization energies (AEs), ionization potentials (IPs) and electron
        affinities (EAs) as depicted in Fig.~\ref{fig:aeipea}.
        The mean absolute deviations (MAD), root mean squared (RMS) and maximum deviations (MAX)
        have been computed for basis set incompleteness errors of
        correlation energy contributions to the AEs, IPs and EAs only.
        The CBS limit reference results have been approximated using CCSD/[45].
        F12 results have been taken from Ref.~\cite{Knizia2009}.
\label{tab:aeipea}
}
\begin{ruledtabular}
\begin{tabular}{l ccc|ccc|ccc}
        & \multicolumn{3}{c}{AEs (kJ/mol)}       & \multicolumn{3}{c}{IPs (meV)}  & \multicolumn{3}{c}{EAs (meV)} \\
	& MAD   &   RMS  & MAX                          & MAD & RMS & MAX       & MAD & RMS & MAX\\ \hline
CCSD/AVDZ             &  47.729 & 54.503 & 161.095  & 261.93 & 281.35 & 491.35 & 196.07 & 252.97 & 892.96 \\
CCSD/AVTZ             &  19.972 & 22.046 &  41.375  &  93.85 & 105.82 & 189.12 &  65.20 &  76.31 & 153.93 \\
CCSD/AVQZ             &   7.110 &  8.594 &  17.384  &  37.75 &  42.12 &  76.82 &  24.62 &  28.52 &  52.90 \\
CCSD/AV5Z             &   3.648 &  4.401 &   8.903  &  19.33 &  21.57 &  39.32 &  12.61 &  14.60 &  27.07 \\
\\
CCSD-PPL/AVDZ         &   6.500 & 7.934 & 22.225    & 35.37  & 47.74  & 126.31 & 30.06 & 37.07 & 72.05 \\
CCSD-PPL/AVTZ         &   1.852 & 2.309 &  6.196    & 13.27  & 16.66  &  40.71 & 12.42 & 14.56 & 30.88 \\
CCSD-PPL/AVQZ         &   0.559 & 0.730 &  2.106    &  6.07  &  7.19  &  17.63 &  8.79 &  9.71 & 21.74 \\
\\
CCSD-F12a/AVDZ\cite{Knizia2009}   &  5.972 & 7.031 & 18.202 & 39.02 & 51.96 & 143.45 & 26.79 & 34.93 & 98.90 \\
CCSD-F12a/AVTZ\cite{Knizia2009}   &  1.534 & 1.859 &  4.102 &  7.70 &  9.23 &  23.97 & 10.28 & 11.76 & 18.14 \\
CCSD-F12a/AVQZ\cite{Knizia2009}   &  1.910 & 2.167 &  4.654 &  8.73 &  9.10 &  14.35 & 10.37 & 11.38 & 18.08 \\
\\
CCSD-F12b/AVDZ\cite{Knizia2009}   &  8.390 & 10.230 & 27.434 & 60.47 & 70.84 & 176.81 & 39.24 & 50.22 & 132.47 \\
CCSD-F12b/AVTZ\cite{Knizia2009}   &  1.717 &  2.144 &  6.138 & 21.80 & 44.69 &  44.69 & 11.92 & 14.83 &  30.35 \\
CCSD-F12b/AVQZ\cite{Knizia2009}   &  0.554 &  0.701 &  1.714 &  4.24 &  5.20 &  12.07 &  3.15 &  3.76 &   8.13 \\

\end{tabular}
\end{ruledtabular}
\end{table*}

\begin{figure*}[ht]
\includegraphics[width=18.0cm,clip=true]{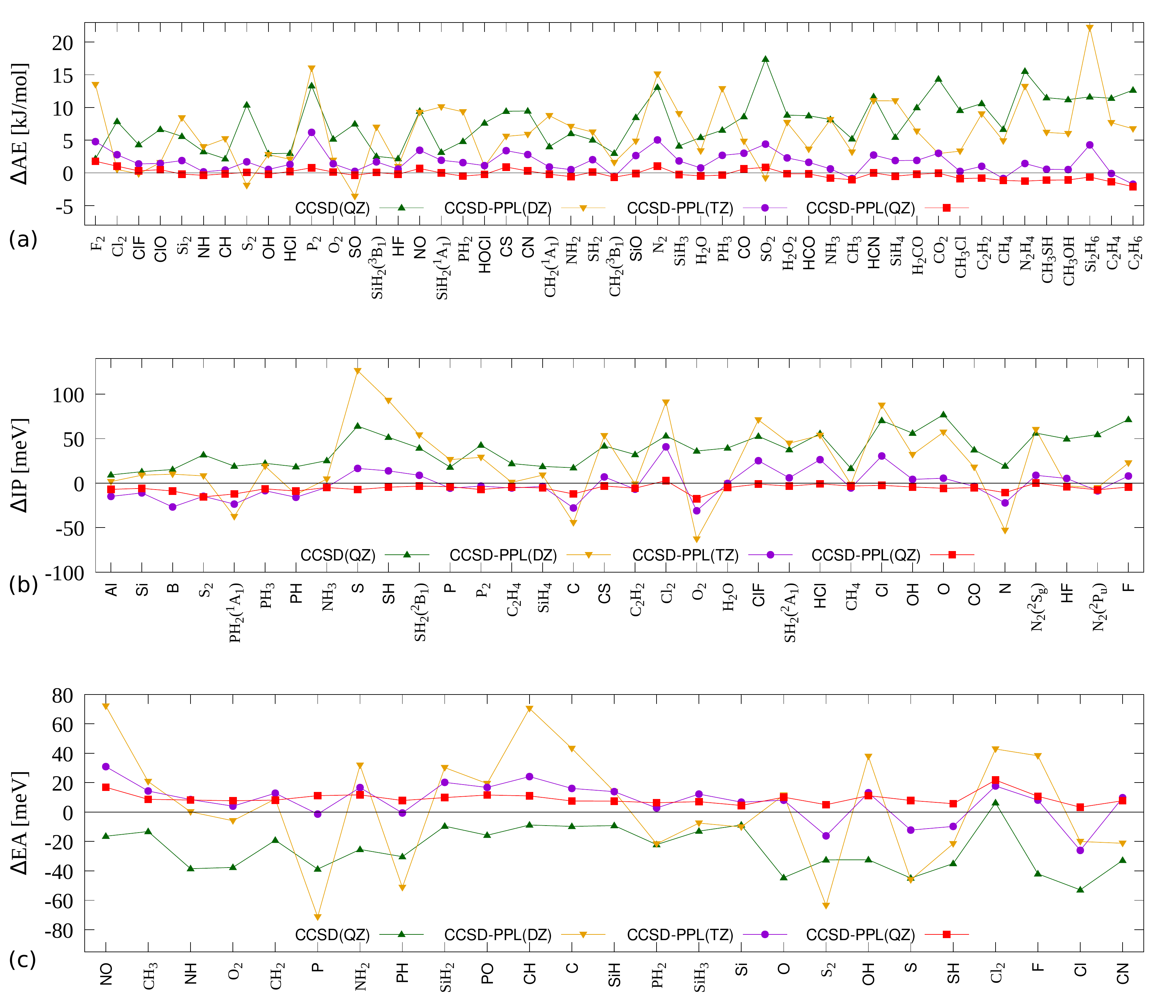}
\caption{\label{fig:aeipea}
	Basis set incompleteness errors of computed correlation energy contributions to
        atomization energies (AEs), ionization potentials (IPs) and electron
        affinities (EAs). A statistical analysis of the errors is summarized in Table~\ref{tab:aeipea}.
        The required CBS limit results have been approximated using CCSD/[45].
}
\end{figure*}

As a further test, we have calculated 
atomization energies (AEs), ionization potentials (IPs), and electron affinities (EAs)
using CCSD and CCSD-PPL. Figure~\ref{fig:aeipea} depicts the CBS error in the valence
electron correlation energy contributions.

For the atomization energies shown in Fig.~\ref{fig:aeipea} (top panel), we observe a
monotonically decreasing error for increasing basis set size. CCSD-PPL/AVDZ achieves
on average the same level of precision as CCSD/AVQZ. The same conclusion can be drawn from the
statistical analysis given in Table~\ref{tab:aeipea}. It is shown that not only mean absolute
and root mean squared deviations are reduced, but also the maximum deviation.
A comparison of CCSD-PPL to CCSD-F12a reveals that on average CCSD-F12a achieves
a slightly better agreement with CBS limit values for AVDZ and AVTZ.
For the AVQZ basis set, CCSD-PPL is closer to the CBS limit estimates
than CCSD-F12a and shows comparable deviations as CCSD-F12b.

In the case of the basis set errors of the ionization potentials shown in Fig.~\ref{fig:aeipea} (middle panel),
and summarized in Table~\ref{tab:aeipea}, a slightly different situation
arises. Although both the MAD and RMS deviations of CCSD/AVQZ and CCSD-PPL/AVDZ are comparable,
CCSD-PPL shows a few outliers which are significantly larger in magnitude. However, this
behavior is no more present for CCSD-PPL/AVTZ and CCSD/AV5Z,
where not only mean absolute deviations but also the maximum deviations are comparable.
The comparison of CCSD-PPL to F12 methods shows
superior results of F12a for AVDZ and AVTZ basis sets, and better values for F12b for AVQZ basis sets.

Similar conclusions can be drawn for the basis set errors of electron affinities depicted in
Fig.~\ref{fig:aeipea} (bottom panel) and listed in Table~\ref{tab:aeipea}.
Here CCSD-PPL/AVDZ shows
larger maximum and mean deviations than CCSD/AVQZ. However, CCSD-PPL/AVTZ and CCSD/AV5Z yield
very similar results, indicating that the PPL scheme can reduce the required basis set size by two
cardinal numbers compared to CCSD theory.
The comparison with F12 methods leads to the same conclusions as for the ionization potentials.

%
%
%

\subsection{Further analysis}

\begin{table*}
\caption{
	RMS deviations of basis set incompleteness error for REs, AEs, IPs, and EAs. 
        Shown are the MP2 term, the ppl contributions, and the
        energy contributions from the terms denoted as rest. Reference energy of the 
        rest energy is AV5Z, whereas the other two references are obtained from 
        [45]-extrapolation, respectively. CCSD results  
        from Tables~\ref{tab:statreactions} and \ref{tab:aeipea} are presented once  
        more for the sake of readability.
}
\label{tab:rest}
\begin{ruledtabular}
\begin{tabular}{l cccc }
	 & REs (kJ/mol) & AEs (kJ/mol) & IPs (meV) & EAs (meV) \\ \hline
CCSD/AVDZ    & 13.837  & 54.503  & 281.35  & 252.97  \\
CCSD/AVTZ    &  6.485  & 22.046  & 105.82  &  76.31  \\
CCSD/AVQZ    &  2.545  &  8.594  &  42.12  &  28.52  \\
CCSD/AV5Z    &  1.305  &  4.401  &  21.57  &  14.60  \\
\\
$\Delta E^{\rm{rest}}$/AVDZ & 3.230 & 15.199 & 92.60 & 63.87 \\
$\Delta E^{\rm{rest}}$/AVTZ & 1.726 &  4.834 & 30.54 & 25.07 \\
$\Delta E^{\rm{rest}}$/AVQZ & 0.360 &  0.789 &  3.35 &  2.63 \\
\\
$\Delta E^{\rm{driver}}$/AVDZ & 14.396 & 71.041 & 298.31 & 349.61 \\
$\Delta E^{\rm{driver}}$/AVTZ &  6.457 & 26.294 & 126.15 & 106.33 \\
$\Delta E^{\rm{driver}}$/AVQZ &  2.719 & 11.370 &  59.38 &  49.51 \\
$\Delta E^{\rm{driver}}$/AV5Z &  1.391 &  5.823 &  30.40 &  25.35 \\
\\
$\Delta E^{\rm{ppl}}$/AVDZ & 2.201 & 24.111 & 111.64 & 135.07 \\
$\Delta E^{\rm{ppl}}$/AVTZ & 1.616 &  9.864 &  54.45 &  54.14 \\
$\Delta E^{\rm{ppl}}$/AVQZ & 0.906 &  4.397 &  26.16 &  25.30 \\
$\Delta E^{\rm{ppl}}$/AV5Z & 0.464 &  2.252 &  13.40 &  12.95 \\
\end{tabular}
\end{ruledtabular}
\end{table*}

There are significant differences in the CCSD-PPL basis-set convergence between 
the reaction energies on the one hand and atomization energies, ionization potentials,
and electron affinites on the other hand.
As mentioned before, the two presented F12 variants achieve a reduction of the basis set incompleteness 
error by two cardinal numbers for all explored quantities. A similar reduction is achieved using CCSD-PPL
for AEs, IPs and EAs. However, in the case of REs, CCSD-PPL achieves a reduction by one cardinal number only.
In order to investigate this disparity further, we perform an analysis of the
channel decomposed basis set error, summarized in Table~\ref{tab:rest}.
We first investigate the error for $\Delta E^{\mathrm rest}$, which corresponds to the basis set error of
the hypothetically perfect CCSD-PPL scheme assuming driver and ppl contributions in the CBS limit.
We find that the basis set convergence of $\Delta E^{\mathrm rest}$ is much better balanced for all
investigated properties, achieving a reduction by almost two cardinal numbers.
In particular $\Delta E^{\mathrm rest}/AVTZ$ is very close to $CCSD/AV5Z$ for AEs, REs and IPs.
We note that the convergence for EAs is slightly slower, which agrees with findings
from Ref.~\cite{Knizia2009} that have been obtained using explicitly correlated methods and show that
EAs converge more rapidly if additional diffuse basis functions are employed.

The above discussion shows that the employed approximation to $\Delta E^{\rm{ppl}}$
as defined by Eq.~(\ref{eq:DPPL}) is less reliable for REs and achieves results
for AEs, IPs and EAs that are partly fortuitously close to the CBS limit.
To better understand the disparity between REs and the other properties,
Table~\ref{tab:rest} also lists the RMS basis set errors of the particle-particle
ladder and driver terms for all employed basis sets.
We find that the ratio between driver and ppl RMS basis set errors
is larger for REs compared to the AEs, IPs and EAs especially when using small basis sets such as AVDZ and AVTZ.
Furthermore Table~\ref{tab:rest} shows that the basis set error of the ppl contributions to the REs are smaller than for AEs.
Consequently, the simple rescaling procedure defined by Eq.~(\ref{eq:DPPL}) is likely
to introduce non-negligible errors for the reaction energies.

In passing we also note that the basis set errors of the combined rest and ppl contributions partly cancel
(not shown explicitly in Table~\ref{tab:rest}), which is in agreement with findings by Ranasinghe and 
Petersson in their CCSD studies~\cite{Ranasinghe2013}. 


\section{\label{sec:conclusion}Summary and Conclusions}

We have applied a recently introduced finite basis-set correction
method for coupled cluster singles and doubles theory to atoms and molecules.
The method is based on a decomposed correlation energy and
corrects for the basis-set incompleteness error in the second-order
and the particle-particle ladder term of the CCSD energy only.
Compared to CCSD, CCSD-PPL allows for a reduction of atom centered correlation
consistent Gaussian basis sets by approximately two cardinal numbers, while
capturing a similar fraction of the CBS limit correlation energy contributions to
total energies, atomization energies, ionization potentials and electron affinities.
We have also studied reaction energies using CCSD-PPL and CCSD, finding
that the PPL scheme allows for reductions in the  basis set size by one cardinal number only.
In passing we note that F12 theories achieve a reduction by two cardinal numbers for
all properties studied in this work including reaction energies.
We have shown that the discrepancy in the CCSD-PPL basis set convergence between reaction energies
and the other properties can be attributed to the CBS limit approximations to
the particle-particle ladder term.
Nonetheless, we conclude that CCSD-PPL exhibits a good trade-off between
computational cost and basis set reduction for CCSD calculations
in atoms, molecules and solids.
Future work will focus on more accurate and equally efficient approximations to this term.


We stress that it is possible to transfer the outlined method to
other widely-used many-electron theories that allow for a decomposition of
the electronic correlation energy 
such as full configuration interaction and related theories.
Finally, we note that the employed decomposition of the electronic correlation energy
will potentially also be useful for the further development and improvement
of correlation factors used in F12 theories.

\section{Acknowledgements}
Support and funding from the European
Research Council (ERC) under the European Unions
Horizon 2020 research and innovation program (Grant Agreement No 715594) is gratefully
acknowledged.
A. G. gratefully acknowledges many helpful discussions on F12 theory with
Seiichiro Ten-no and David Tew.



\bibliography{article}

\end{document}